\documentclass{iopart}

\usepackage{iopams}
\usepackage{setstack}
\usepackage{graphicx}
\usepackage{bm}

\newcommand{\eqref}[1]{(\ref{#1})}

\newcommand{\eq}[1]{\begin{equation}#1\end{equation}}
\newcommand{\eqarray}[1]{\begin{eqnarray}#1\end{eqnarray}}

\newcommand{\ket}[1]{\ensuremath{\,|\,{#1}\,\rangle}}

\newcommand{\matrixe}[3]{\ensuremath{\langle\,{#1}\,|\,{#2}\,|\,{#3}\,\rangle}}
\newcommand{\op}[1]{\ensuremath{\bm{#1}}}
\newcommand{\dd}{\ensuremath{\mathrm{d}}}

\newcommand{\intd}[2][]{\ensuremath{ \int\!\!\mathrm{d}^{#1}#2\;}}

\newcommand{\vO}{\ensuremath{\op{v}}}

\newcommand{\HO}{\ensuremath{\op{H}}}

\newcommand{\kV}{\ensuremath{\vec{k}}}

\newcommand{\qV}{\ensuremath{\vec{q}}}

\newcommand{\xV}{\ensuremath{\vec{x}}}

\newcommand{\pOV}{\ensuremath{\vec{\op{p}}}}
\newcommand{\qOV}{\ensuremath{\vec{\op{q}}}}
\newcommand{\rOV}{\ensuremath{\vec{\op{r}}}}
\newcommand{\rON}{\ensuremath{\hat{\op{r}}}}

\newcommand{\EC}{\ensuremath{\mathcal{E}}}
\newcommand{\FC}{\ensuremath{\mathcal{F}}}

\begin{document}

\title[Phase diagram of trapped degenerate Fermi gases]
  {Phase diagram of trapped degenerate Fermi gases
  including effective s- and p-wave interactions}

\author{R. Roth and H. Feldmeier}

\address{Gesellschaft f\"ur Schwerionenforschung (GSI)\\
  Planckstr. 1, 64291 Darmstadt, Germany}

\ead{\mailto{r.roth@gsi.de}, \mailto{h.feldmeier@gsi.de}}

\begin{abstract}
The influence of s- and p-wave interactions on trapped degenerate one
and two-component Fermi gases is investigated. The energy functional
of a multicomponent Fermi gas is derived within the Thomas-Fermi
approximation including the s- and p-wave terms of an effective
contact interaction. On this basis the stability of the dilute gas
against mean-field induced collapse due to attractive interactions is
investigated and explicit stability conditions in terms of Fermi
momentum and s- and p-wave scattering lengths are derived. Furthermore
the spatial separation in a two-component Fermi gas is discussed,
explicit conditions for the onset of component demixing are given, and
the density distributions in the separated phase are calculated. The
findings are summarized in a zero-temperature phase diagram for the
degenerate two-component Fermi gas. It is shown that the p-wave
interaction has significant influence on the phase diagram of the
degenerate Fermi gas and causes new phenomena like absolute
stabilization against collapse and component separation. It may
therefore be useful in the context of the envisioned BCS transition in
trapped atomic Fermi gases.
\end{abstract}




\section{Introduction}

Looking back at the outstanding experimental achievements in the field
of trapped ultracold gases composed of bosonic atoms of the last 6
years one wonders what exciting new developments one has to envision in
the near future \cite{ProcEF99}. One branch of these developments
will involve dilute and degenerate gases of fermionic atoms. Already
today several groups \cite{DeJi99,TrSt01,ScKh01} were able to cool
samples of fermionic atoms, ${}^6$Li or ${}^{40}$K, to temperatures
corresponding to a fraction of the Fermi energy. These experiments use
either a single fermionic isotope (${}^{40}$K) in two different
hyperfine states \cite{DeJi99} or a mixture of a fermionic and a
bosonic species (${}^6$Li/${}^7$Li) \cite{TrSt01,ScKh01}. In this way
s-wave scatterings between atoms of either the two fermionic
components or between bosons and fermions allow efficient evaporative
cooling of the gas. Remember that in a gas of identical fermions
s-wave interactions are suppressed by the Pauli principle.

The experiments on dilute degenerate Fermi gases open a new and
unique field to study fundamental quantum phenomena. Many of the
questions discussed for the Bose-Einstein condensed Bose gas also apply
to degenerate Fermi gases. Examples are the mean-field induced
collapse of the dilute gas due to attractive interactions
\cite{DoCl01} or the variety of collective excitations \cite{OnDu00}.
One of the most challenging experimental goals is the realization of a
BCS-transition from the normal to a superfluid phase
\cite{HoFe97,StHo99}. However, a detailed understanding of the
``normal'' degenerate Fermi gas is a prerequisite to achieve that.

In this paper we investigate the influence of interactions on
single-component and two-component Fermi gases in the normal phase
with special emphasis on the role of the p-wave interaction. Therefore
in section \ref{sec:ef} we construct an effective s- and p-wave
contact interaction and use it to derive the energy density of a
multicomponent Fermi gas in Thomas-Fermi approximation. On this basis
we investigate the structure of one- and two-component Fermi gases at
zero temperature. A first question concerns is the stability of the
dilute gas against collapse induced by attractive interactions. In
section \ref{sec:col} we derive stability conditions for the one- and
two-component gas and discuss the influence of the p-wave
interaction. A second question deals with the spatial separation of
the different components in a two-component system. Conditions for the
onset of component demixing are derived in section \ref{sec:sep}. We
also discuss the structure of the density distributions in the
separated phase. Finally, in section \ref{sec:phasediag} we summarize
our findings by means of a zero-temperature phase diagram for the
two-component Fermi gas.

\section{Energy functional of a multicomponent Fermi gas}
\label{sec:ef}

\subsection{Mean-field approximation and effective contact interaction}

In general the description of an interacting quantum mechanical
many-body system relies on approximations to render the problem
solvable. A general procedure to obtain different approximation
schemes is to restrict the many-body states the system can assume to a
certain subspace---called model space---of the full Hilbert
space. A prominent example is the mean-field approximation where the
many-body state is given by a symmetrized or antisymmetrized product
of single-particle states. Within this mean-field model space the
many-body problem can be solved; in case of fermions the solution is
obtained by the well known Hartree-Fock equations.

However, this restriction to a model space of sufficiently simple
structure encounters a serious drawback. Many realistic interactions
between the constituents of the many-body system exhibit a strongly
repulsive short-range core in addition to the attractive part.
Typical examples are van der Waals interactions between neutral atoms,
e.g. the Lennard-Jones potential with an attractive part
proportional to $r^{-6}$ and a strong repulsion proportional to
$r^{-12}$.  The repulsive core prevents any of the particles to
approach each other closer than the core radius, in other words, the
two-body density distribution is depleted or even zero for particle
distances smaller than the core radius. These short-range correlations
cannot be described within a simple model space like the one of the
mean-field approximation \cite{RoFe01b,FeNe98}.

To resolve this discrepancy we transform the realistic potential into
an effective interaction suited for the particular model space under
consideration. In the case of cold and dilute atomic gases the
properties of the physical system simplify the construction of the
effective interaction significantly. This is our starting point: (i)
We want to describe the properties of a degenerate Fermi gas in
the mean-field approximation, i.e. the many-body state is an
antisymmetrized product state. (ii) The system is cold and dilute,
i.e. mean particle distance and wavelength of relative motion are much
larger than the range of the two-body potential. (iii) The system is
in a non self-bound metastable many-body state, the two-body bound
states of the potential are not populated.

Due to (ii) and (iii) the atoms do not probe the detailed radial
dependence of the interaction, therefore we can use a simple
zero-range effective potential. To describe interactions also between
particles with relative angular momenta $l>0$ the contact interaction
has to be nonlocal.  The mapping of the exact onto the effective
interaction is done by means of the two-body energy spectrum: We
require that the exact two-body eigenvalues $E_{nl}^{\mathrm{exact}}$
for positive energies (scattering states) is reproduced by the
expectation values of the effective interaction calculated with
kinetic energy eigenstates $\ket{nlm_l}$
\eq{ \label{eq:ef_eci_condition}
  \matrixe{nlm_l}{\case{1}{2m_{\mathrm{red}}}\qOV^2 +
 \vO^{\mathrm{ECI}}}{nlm_l}
  \overset{!}{=} E_{nl}^{\mathrm{exact}}
}
where $\qOV = \frac{1}{2}(\pOV_1-\pOV_2)$ is the relative momentum of
the two particles and $m_{\mathrm{red}}$ their reduced mass. In the
following we will consider only particles with equal masses,
i.e. $m_{\mathrm{red}}=m/2$. $l$ and $m_l$ are the quantum numbers of
relative angular momentum, and $n$ is a radial quantum number
\cite{RoFe01a}.  Guided by \eqref{eq:ef_eci_condition} we can
construct the operator of the effective contact interaction (ECI)
explicitly. Up to the quadratic order in the relative momentum $\qV$,
which includes a p-wave contribution, the operator of the ECI reads
\cite{RoFe01a}
\eqarray{ \label{eq:ef_eci_operator} \fl
  \vO^{\mathrm{ECI}} 
  = \frac{4\pi\,a_0}{m}\, \delta^{(3)}(\rOV)
  + \frac{6\pi\,b_0}{m} [\delta^{(3)}(\rOV)\, (\rON\cdot\qOV)^2 
    + \mathrm{h.a.}] 
  + \frac{12\pi\,a_1^3}{m}\,(\qOV\cdot\rON)\,
    \delta^{(3)}(\rOV)\,(\rON\cdot\qOV) ,
}
where $\rON=\rOV/|\rOV|$ and $\hbar=1$. The properties of the exact
two-body interaction enter through the s-wave scattering length $a_0$,
the s-wave effective volume $b_0$, and the p-wave scattering length
$a_1$. They are defined by the first terms of the low-momentum
expansion of the scattering phase shifts $\eta_l(q)$ of the exact
potential
\eqarray{
  \frac{\tan\eta_0(q)}{q} \approx -a_0 - b_0 q^2  
  \;,\qquad
  \frac{\tan\eta_1(q)}{q^3} \approx -\case{1}{3}a_1^3 .
}

The first term of the effective contact interaction
\eqref{eq:ef_eci_operator} is identical to the widely used s-wave
pseudopotential \cite{HuYa57,DaGi99}. The two additional nonlocal
terms are the natural extension to describe effective range effects
and more important the effects of p-wave interactions.

One should be aware that the effective interaction
\eqref{eq:ef_eci_operator} is designed for the application in a
mean-field model space. It does not make sense to use this interaction
in a larger model space or even for the solution of the Schr\"odinger
equation \cite{RoFe01a}. When using erroneously this contact
interaction to solve the exact scattering problem one does not get
back the correct phase shifts.

\subsection{Energy density of a multicomponent Fermi gas}

We construct the energy density of the inhomogeneous Fermi gas in an
external trapping potential within the Thomas-Fermi approximation,
which was shown \cite{Roth00} to be an excellent approximation for the
particle numbers considered here. The basic assumption of the
Thomas-Fermi approximation is that the energy density of the
inhomogeneous system is given locally by the energy density of the
corresponding homogeneous system.

The calculation of the energy density of the homogeneous
multicomponent Fermi gas in mean-field approximation is
straightforward. The translational invariant part of the Hamiltonian
of the trapped gas, consists of the kinetic energy operator and the
effective contact interaction \eqref{eq:ef_eci_operator}
\eq{ \label{eq:ef_hamiltonian_int} 
  \HO_{\mathrm{hom}} 
  = \frac{1}{2m} \sum_i^N \pOV_i^2  
  + \sum_{i,\,j>i}^N \vO^{\mathrm{ECI}}_{ij} . 
}
The energy density in mean-field approximation is defined by the
expectation value of this Hamiltonian calculated with a Slater
determinant $\ket{\Psi}$ 
\eq{  \label{eq:ef_energydens_hom_def}
  \EC_{\mathrm{hom}} 
  = \frac{1}{V} \matrixe{\Psi}{\HO_{\mathrm{hom}}}{\Psi} .
}
The single-particles states $\ket{i}$ are eigenstates of the momentum
operator with momenta $\kV_i$ up to the Fermi momentum $\kappa$. To
account for the different $m_F$ substates that can be trapped
simultaneously \cite{DeJi99} we characterize the single-particle
states by an additional component quantum number $\xi=1,\dots,\Xi$
\eq{
  \ket{i} 
  = \ket{\kV_i, \xi_i} .
}
The many-body state $\ket{\Psi}$ is thus an antisymmetrized product of
all single-particle states $\ket{i}$ up to the Fermi momenta
$\kappa_{\xi}$ of the different components $\xi$. The calculation of
the expectation value \eqref{eq:ef_energydens_hom_def} assuming a
cubic box of volume $V$ with periodic boundary conditions is
straightforward.  Within the Thomas-Fermi approximation
$\EC_{\mathrm{hom}}$ can be directly translated into the energy
density of the inhomogeneous trapped system by replacing the Fermi
momenta $\kappa_{\xi}$ with \emph{local} Fermi momenta
$\kappa_{\xi}(\xV)$ and adding the contribution of the trapping
potential $U_{\xi}(\xV)$. This leads to the general form of the energy
density of a trapped multicomponent Fermi gas including s- and p-wave
interactions \cite{RoFe01a,RoFe00}
\eqarray{ \label{eq:ef_energydens} \fl
  \EC[\kappa_1,\dots,\kappa_{\Xi}](\xV) 
  = \frac{1}{6\pi^2}  \sum_{\xi} U_{\xi}(\xV)\; \kappa_{\xi}^3(\xV) 
    + \frac{1}{20 \pi^2 m}  \sum_{\xi} \kappa_{\xi}^5(\xV) 
    + \frac{a_0}{9 \pi^3 m}  \sum_{\xi,\,\xi'>\xi}
    \kappa_{\xi}^3(\xV) \kappa_{\xi'}^3(\xV)     \nonumber\\
  \lo+ \frac{a_1^3}{30 \pi^3 m}  \sum_{\xi} \kappa_{\xi}^8(\xV) 
    + \frac{a_1^3+b_0}{60 \pi^3 m}  \sum_{\xi,\,\xi'>\xi} 
    [ \kappa_{\xi}^3(\xV) \kappa_{\xi'}^5(\xV) 
    + \kappa_{\xi}^5(\xV) \kappa_{\xi'}^3(\xV) ]  , 
}
where the summations range over all components
$\xi,\xi'=1,\dots,\Xi$. The interaction is characterized by the s-wave
scattering length $a_0$, the s-wave effective volume $b_0$, and the
p-wave scattering length $a_1$. In order to keep the treatment as
transparent as possible we have assumed equal masses $m$ and
interaction parameters for the atoms of all components. The local
Fermi momentum $\kappa_{\xi}(\xV)$ is connected to the partial density
$\rho_{\xi}(\xV)$ of the respective component by
\eq{
  \rho_{\xi}(\xV) = \frac{1}{6\pi^2}\,\kappa_{\xi}^3(\xV) .
}
For most of the following investigations the detailed form of the
trapping potential $U_{\xi}(\xV)$ does not enter. If, however, the
shape of the trap is relevant, then we assume a deformed
harmonic-oscillator potential
\eq{ \label{eq:ef_trappingpot}
  U(\xV) 
  = \frac{m}{2} \big(\omega_1^2 x_1^2 
    + \omega_2^2 x_2^2 + \omega_3^2 x_3^2 \big)\; ,
  \qquad
  \xV=(x_1,x_2,x_3)\;,  
}
with a mean oscillator frequency
$\omega=\sqrt[3]{\omega_1\omega_2\omega_3}$ and the corresponding
mean oscillator length $\ell = (m\omega)^{-1/2}$.

\subsection{Functional variation and extremum condition}

On the basis of the energy density functional \eqref{eq:ef_energydens}
we obtain the density distribution of the degenerate Fermi gas by
functional minimization of the energy
\eq{
  E[\kappa_1,\dots,\kappa_{\Xi}] 
  = \int\!\!\dd^3x\; \EC[\kappa_1,\dots,\kappa_{\Xi}](\xV) .
}
with the constraints of given numbers of atoms $N_{\xi}$ of the
different components $\xi = 1,\dots,\Xi$
\eq{
  N_{\xi} 
  = \intd[3]{x} \rho_{\xi}(\xV)
  = \frac{1}{6\pi^2} \intd[3]{x} \kappa^3_{\xi}(\xV) .
}
For the variational treatment these constraints are implemented via a
set of Lagrange multipliers $\mu_{\xi}$---the chemical potentials of
the different components. This leads to the Legendre transformed
energy
\eq{ 
  F[\kappa_1,\dots,\kappa_{\Xi}] 
  = \int\!\!\dd^3x\; \FC[\kappa_1,\dots,\kappa_{\Xi}](\xV) ,
}
where the associated energy density
$\FC[\kappa_1,\dots,\kappa_{\Xi}](\xV)$ is given by
\eq{ \label{eq:ef_transf_energydens}
  \FC[\kappa_1,\dots,\kappa_{\Xi}](\xV) 
  = \EC[\kappa_1,\dots,\kappa_{\Xi}](\xV) 
    - \sum_{\xi}\frac{\mu_{\xi}}{6\pi^2}\kappa_{\xi}^3(\xV) . 
}
The set of density distributions $\rho_{\xi}(\xV)$ or local Fermi
momenta $\kappa_{\xi}(\xV)$ that minimizes the transformed energy
describes the state of the trapped degenerate Fermi gas. A necessary
condition for a minimum of the energy $F[\kappa_1,\dots,\kappa_{\Xi}]$
is that the derivative of the energy density
$\FC[\kappa_1,\dots,\kappa_{\Xi}](\xV)$ with respect to the local
Fermi momentum vanishes for this set
\eq{
  \frac{\partial}{\partial\kappa_{\xi}(\xV)} 
  \FC[\kappa_1,\dots,\kappa_{\Xi}](\xV) = 0
  \qquad\mathrm{for\; all}\quad\xV,\,\xi .
}
This condition is also fulfilled by maxima or saddle points, therefore
one has to check explicitly whether the solutions describe a minimum
of $F[\kappa_1,\dots,\kappa_{\Xi}]$. Inserting the energy density
\eqref{eq:ef_transf_energydens} leads to a set of $\Xi$ coupled
polynomial equations for the local Fermi momenta at the point $\xV$
\eqarray{ \label{eq:ef_extremumcond} \fl 
  m[\mu_{\xi}-U_{\xi}(\xV)]
  = \frac{1}{2}\kappa_{\xi}^2(\xV) 
    + \frac{2 a_0}{3\pi} \sum_{\xi'\ne\xi} \kappa_{\xi'}^3(\xV)  
    + \frac{8 a_1^3}{15\pi} \kappa_{\xi}^5(\xV)  \nonumber\\
  \lo+ \frac{a_1^3+b_0}{30\pi} \sum_{\xi'\ne\xi} [ 3\,\kappa_{\xi'}^5(\xV) 
    + 5\,\kappa_{\xi}^2(\xV)\,\kappa_{\xi'}^3(\xV) ] 
  \;,\qquad \xi = 1,\dots,\Xi .
}
These extremum conditions are the starting point for the investigations
on the stability of the trapped Fermi gas presented in the following
sections.

\section{Mean-field collapse}
\label{sec:col}

In the presence of attractive interactions between the trapped atoms,
i.e., if the (s- or p-wave) scattering length is negative, the dilute
gas may undergo a collective collapse towards a high-density
configuration. The physical origin of the collapse is the attractive
mean-field generated by the interaction. At sufficiently low densities
this mean-field attraction is compensated by the positive kinetic
energy contribution. However, if the density grows the negative
mean-field contribution to the energy grows faster than the kinetic
energy. At some critical density the kinetic part is not able to
stabilize the gas against the mean-field attraction and the system
collapses. For bosonic systems this phenomenon was observed in a
series of outstanding experiments \cite{RoCl01}.

The same phenomenon can occur in degenerate Fermi gases with attractive
s- or p-wave interactions. In the following we derive explicit
conditions for the stability of one- and two-component systems.

\subsection{Stability of the single-component Fermi gas}
\label{sec:col_1comp}

For the single-component system the Legendre transformed energy
density \eqref{eq:ef_transf_energydens} of the multicomponent Fermi
gas reduces to
\eq{ \label{eq:col_1comp_transf_energydens} 
  \FC_1[\kappa](\xV)
  = \frac{1}{6\pi^2} [U(\xV)-\mu] \kappa^3(\xV)
  + \frac{1}{20\pi^2 m} \kappa^5(\xV)
  + \frac{a_1^3}{30\pi^3 m} \kappa^8(\xV) .
}
As a consequence of the Pauli principle the s-wave contact interaction
does not contribute in a system of identical fermions. However, the
p-wave contact interaction does contribute and constitutes the leading
interaction term in the single-component gas.

\begin{figure}
\begin{indented}\item[]\hskip-0.08\textwidth
\includegraphics[width=0.55\textwidth]{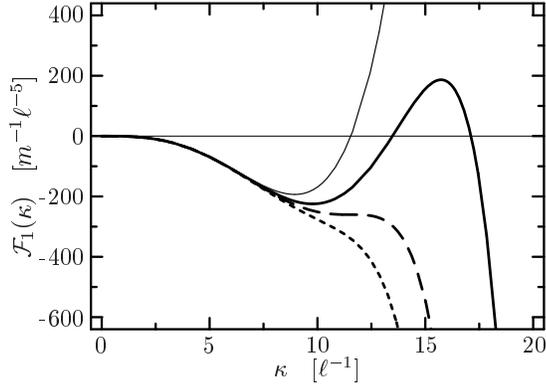}
\end{indented}
\vskip-2.3ex
\caption{Legendre transformed energy density $\FC_1(\kappa)$ of the
  single-component Fermi gas as function of the Fermi momentum
  $\kappa$ for $m\ell^2[\mu-U(\xV)]=40$. The different curves
  correspond to different p-wave scattering lengths: $a_1/\ell=0$
  (thin solid line), $-0.08$ (solid), $-0.0915$ (dashed), and $-0.1$
  (dotted). }
\label{fig:col_1comp_energy}
\end{figure}

To illustrate the origin of the mean-field instability figure
\ref{fig:col_1comp_energy} shows the energy density
\eqref{eq:col_1comp_transf_energydens} in the centre of the trap as
function of the Fermi momentum $\kappa$ for a fixed value of the
chemical potential $\mu$ and different p-wave scattering lengths. We
express all quantities in units of the atom mass $m$ and a length
scale $\ell$, which later on will be the mean oscillator length of the
trapping potential. The thin solid curve shows the energy density of a
noninteracting Fermi gas ($a_1/\ell=0$). The minimum defines the
actual Fermi momentum at the given position $\xV$ according to the
variational description discussed above. If we switch on an attractive
p-wave interaction ($a_1/\ell<0$), then the dominant $\kappa^8$ term
in \eqref{eq:col_1comp_transf_energydens} leads to a rapid decline of
the energy at large Fermi momenta. For sufficiently weak attractions
(thick solid curve) a local minimum at low Fermi momenta
remains. However, if the strength of the p-wave interaction reaches a
critical value the minimum evolves to a saddle point (dashed
curve). Beyond this value of the scattering length the energy density
drops monotonically and there is no low-density solution any more
(dotted curve in figure \ref{fig:col_1comp_energy}). We conclude that
for a given chemical potential there is a limit to the scattering
length up to which a metastable low-density solution exists.

The analytic criterion for the instability of the Fermi gas is the
vanishing of the local minimum at low densities; the onset of
instability is marked by appearance of a the saddle point, where first
and second derivative of the transformed energy density
$\FC_1(\kappa)$ vanish. It is advantageous to perform the derivatives
of \eqref{eq:col_1comp_transf_energydens} with respect to density
$\rho=\kappa^3/(6\pi^2)$ rather than Fermi momentum. This leads to a
condition for the existence of the minimum involving the p-wave
scattering length $a_1<0$ and the local
Fermi momentum $\kappa$
\footnote{Alternatively these conditions can be obtained from the
extremum condition \eqref{eq:ef_extremumcond}. The collapse region is
characterized by parameters where \eqref{eq:ef_extremumcond} has
no real positive solution anymore \cite{RoFe01a,RoFe00}.}:
\eq{ \label{eq:col_1comp_stabilitycond}
  -a_1\,\kappa(\xV) \le \frac{\sqrt[3]{3\pi}}{2} 
  \qquad\mathrm{and}\qquad
  m[\mu-U(\xV)]\, a_1^2 \le \frac{3(3\pi)^{2/3}}{40} .
}
The second inequality is obtained by inserting the Fermi momentum of
the inflection point into the extremum condition
\eqref{eq:ef_extremumcond} to obtain a limit for the chemical
potential. These are stability conditions for the trapped
single-component Fermi gas, in the presence of an attractive p-wave
interaction $a_1<0$; any configuration that violates these condition
will collapse.

We obtain a more intuitive relation by replacing the Fermi momentum
with the total number of atoms $N$ in a trap of given geometry; here
we assume a deformed harmonic-oscillator trap
\eqref{eq:ef_trappingpot} with mean oscillator length $\ell$. For a
given set of p-wave scattering lengths and the corresponding critical
chemical potentials according to \eqref{eq:col_1comp_stabilitycond} we
solve the extremum condition numerically and obtain the critical
particle number. The resulting relation between $a_1$ and the critical
particle number $N_{\mathrm{crit}}$ of a metastable trapped Fermi gas
can be parametrized by the stability condition 
\eq{ \label{eq:col_1comp_stabilitycond_N}
  -2.246\;\Big(\!\sqrt[6]{\!N}\frac{a_1}{\ell}\Big) \le 1,  
}
where the prefactor is adjusted according to the pairs of $a_1$ and
$N_{\mathrm{crit}}$ obtained numerically.

For the single-component Fermi gas the instability induced by
attractive p-wave interactions is no severe limitation in the present
experimental parameter regime. For typical particle numbers of
$N\approx10^6$ and a mean oscillator length of $\ell=1\,\mu$m the p-wave
scattering length would have to be of the order of
$a_1\approx-840\,a_{\mathrm{B}}$ (where $a_{\mathrm{B}}$ is the Bohr
radius) to cause the instability. However, by exploiting a p-wave
Feshbach resonance \cite{Bohn00} these values of the scattering length
seem to be reachable.

\subsection{Stability of the two-component Fermi gas}
\label{sec:col_2comp}

The same analysis can be performed for a two-component Fermi gas. In
order to keep the number of parameters small we assume identical local
Fermi momenta $\kappa(\xV)=\kappa_1(\xV)=\kappa_2(\xV)$ for the two
components. A justification of this assumption follows in section
\ref{sec:sep_2comp_stabcond}. For these systems the transformed energy
density \eqref{eq:ef_transf_energydens} reduces to
\eqarray{ \label{eq:col_2comp_transf_energydens} \fl
  \FC_2[\kappa](\xV)
  = \frac{1}{3\pi^2} [U(\xV)-\mu] \kappa^3(\xV)
  + \frac{1}{10\pi^2 m} \kappa^5(\xV) 
  + \frac{a_0}{9\pi^3 m} \kappa^6(\xV)
  + \frac{a_1^3}{10 \pi^3 m} \kappa^8(\xV) ,
}
where the s-wave effective volume $b_0$ is neglected. For this
discussion it can be included subsequently through a modified p-wave
scattering length $a_1^3 \to a_1^3 + b_0/3$. In contrast to the
single-component Fermi gas s-wave as well as p-wave interactions
contribute. Their interplay leads to a rich variety of phenomena.

As for the single-component system a mean-field instability of the
degenerate gas can emerge as consequence of an attractive
interaction. In the two-component case, however, the s-wave or the
p-wave interaction may be attractive which causes a more complicated
stability condition. Evaluating the Fermi momentum associated with the
saddle point of the energy density
\eqref{eq:col_2comp_transf_energydens} leads to the stability
condition
\eq{ \label{eq:col_2comp_stabilitycond_kF}
-a_0\kappa(\xV) - 2\,[a_1\kappa(\xV)]^3 \le \frac{\pi}{2} .
}
It is convenient to discuss the four possible combinations of signs of
the s-wave scattering length $a_0$ and the p-wave scattering length
$a_1$ separately:
\begin{description}
\item[$a_0>0, a_1>0$:] for purely repulsive interactions the gas is
stable against collapse for any Fermi momentum.
\item[$a_0<0, a_1<0$:] for purely attractive interactions both
interaction parts cooperate and lower the critical Fermi momentum
compared to a pure s-wave or p-wave attraction.
\item[$a_0>0, a_1<0$:] the repulsive s-wave interaction stabilizes the
system against p-wave induced collapse, i.e., the critical Fermi
momentum is increased compared to a pure p-wave attraction.
\item[$a_0<0, a_1>0$:] similar to the previous case the p-wave
repulsion stabilizes the system, i.e. increases the critical Fermi
momentum, compared to the case of a pure s-wave attraction. Moreover,
if the ratio of p-wave and s-wave scattering lengths exceeds a limit
given by
\eq{ \label{eq:col_2comp_nolimitcond}
  \frac{a_1}{|a_0|} \ge \frac{2}{3\pi^{2/3}} \approx 0.311 ,
}
then the system is absolutely stabilized by the p-wave repulsion. That
means, despite of the attractive s-wave interaction the mean-field
collapse does not occur. Notice that the p-wave scattering length
required for this absolute stabilization is only 1/3 of the s-wave
scattering length.
\begin{figure}
\begin{indented}\item[]\hskip-0.085\textwidth
\includegraphics[height=0.28\textheight]{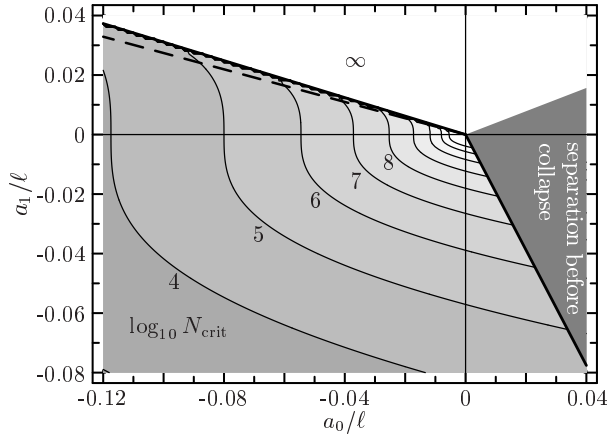}
\end{indented}
\vskip-2.3ex
\caption{Logarithmic contour plot of the critical particle number 
  $N_{\mathrm{crit}}$ up to which the trapped two-component Fermi gas
  is stable against mean-field collapse as function of the ratio of
  s-wave and p-wave scattering lengths to mean oscillator length
  $\ell$ of the parabolic trap. The white area in the upper part
  indicates the parameter region, where no collapse occurs. }
\label{fig:col_2comp_stabmap}
\end{figure}

For repulsive p-wave interactions that are slightly to weak to
generate absolute stabilization, the density profile of the trapped
gas shows a peculiar structure: In a region around the minimum of the
trapping potential the density is increased by typically one order of
magnitude compared to peripheral regions. In contrast to the outer
low-density phase, which is governed by a balance between s-wave
attraction and kinetic energy, the inner high-density phase is ruled
by the balance between s-wave attraction and p-wave repulsion. Without
the p-wave repulsion the high-density phase would collapse
immediately. We discuss this phenomenon in detail in ref. \cite{RoFe01a}.
\end{description}
The last class of interactions (attractive s-wave and repulsive
p-wave) is of special interest in connection with a BCS phase
transition in trapped dilute Fermi gases. The formation of Cooper
pairs of two fermionic atoms requires an attractive interaction. One
possibility is an attractive s-wave interaction in a two-component
Fermi gas which would allow Cooper pairs composed of one atom from
each component. The critical temperature for the BCS transition
increases with the product $|a_0| \kappa$ of scattering length and
Fermi momentum \cite{HoFe97}. At the same time the mechanical
stability of the system is a prerequisite. This sets an upper limit to
the value of $|a_0| \kappa$---and therewith to the BCS transition
temperature---described by the stability condition
\eqref{eq:col_2comp_stabilitycond_kF}. However, if an appropriate
repulsive p-wave interaction is included, then the effect of absolute
stabilization can be employed to get rid of this limitation of the
transition temperature caused by the mean-field instability%
\footnote{Of course it has to be checked that the gain of mechanical
stability due to the p-wave repulsion is not counteracted by a
destructive effect on the Cooper pairs.}.

An overview of the influence of s- and p-wave interactions on the
stability of the two-component Fermi gas against mean-field induced
collapse is given in figure \ref{fig:col_2comp_stabmap}. The contour
plot shows the critical particle number $N_{\mathrm{crit}}$ for a trap
of mean oscillator length $\ell$ as function of the ratios $a_0/\ell$
and $a_1/\ell$. The particle number is calculated numerically in the
same way as described for the single-component gas
\cite{RoFe01a,RoFe00}. The plot confirms the relevance of the p-wave
part of the interaction for the properties---especially for the
stability---of the dilute degenerate Fermi gas. Two major effects of
the p-wave interaction are: (i) Absolute stabilization of the system
by repulsive p-wave interactions in the presence of a s-wave
attraction (white region) and p-wave stabilized high-density phase
(region between the dashed line and the onset of absolute
stabilization). (ii) Mean-field collapse induced by p-wave attraction
also for systems with repulsive s-wave interactions (lower right
quadrant). In addition a competition with another effect, the
component separation that is discussed in the following section,
appears (dark gray area).

As a rule of thumb, the p-wave part becomes relevant as soon as the
p-wave scattering length exceeds about $1/10$ of the value of the
s-wave scattering length. A detailed discussion of the various effects
is presented in \cite{RoFe01a}.

\section{Component separation}
\label{sec:sep}

A second structural transition that can occur in degenerate
two-component Fermi gases is the spatial demixing of the two
components \cite{SaPo00,AmMe00}. The physical mechanism behind this
transition can be understood from the general form of the energy
density of a two-component system resulting from
\eqref{eq:ef_transf_energydens}
\eqarray{ \label{eq:sep_2comp_transf_energydens} \fl
  \FC_2[\kappa_1,\kappa_2](\xV)
  = \frac{1}{6\pi^2} \big[(U_1(\xV)-\mu_1)\, \kappa_1^3(\xV)
    + (U_2(\xV)-\mu_2)\, \kappa_2^3(\xV)\big] \nonumber \\
  \lo+ \frac{1}{20\pi^2 m} \big[\kappa_1^5(\xV) + \kappa_2^5(\xV)\big]
  + \frac{a_0}{9\pi^3 m}\, \kappa_1^3(\xV) \kappa_2^3(\xV) \\
  \lo+ \frac{a_1^3}{30 \pi^3 m} \big[\kappa_1^8(\xV) + \kappa_2^8(\xV)
  + \case12 \kappa_1^3(\xV) \kappa_2^5(\xV)
  + \case12 \kappa_1^5(\xV) \kappa_2^3(\xV) \big] \nonumber.
}
The driving term for component separation is the s-wave part of the
interaction. Its contribution to the energy density is proportional to
the product of the densities of both components
$\rho_1(\xV)\,\rho_2(\xV) \propto \kappa_1^3(\xV)\,\kappa_2^3(\xV)$. For
strongly repulsive s-wave interactions it is energetically favourable
to spatially separate the two components such that the overlap of the
two density distributions and thus the contribution of the s-wave part
to the total energy is reduced.
\begin{figure}
\begin{indented}\item[]\hskip-0.06\textwidth
\includegraphics[width=0.78\textwidth]{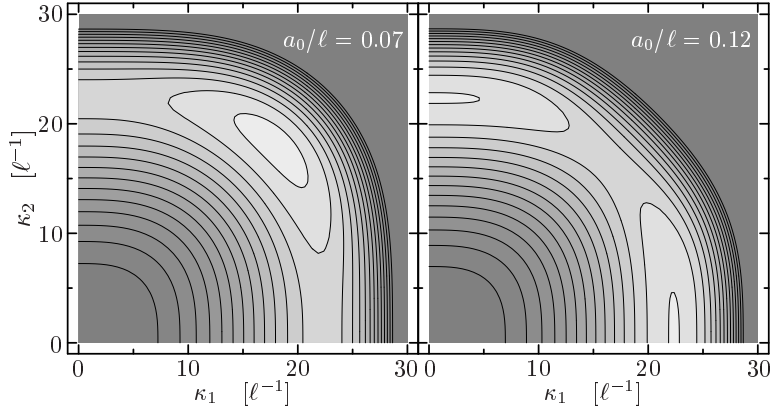}
\end{indented}
\vskip-2.3ex
\caption{Contour plots of the local energy density
  $\FC_2(\kappa_1,\kappa_2)$ of a two-component Fermi gas in the
  centre of the trap as function of the Fermi momenta $\kappa_1$ and
  $\kappa_2$ for $\mu_1=\mu_2=250\,m^{-1}\ell^{-2}$. The left panel
  shows the case of a moderate s-wave repulsion ($a_0/\ell=0.07,\,
  a_1/\ell=0$) that still supports identical density distributions for both
  components. The right panel corresponds to a stronger s-wave
  repulsion ($a_0/\ell=0.12,\,a_1/\ell=0$) which forces the components to
  demix partially. Brighter shading corresponds to smaller values of
  the energy density.}
\label{fig:sep_2comp_energy}
\end{figure}

A quantitative example is shown in figure
\ref{fig:sep_2comp_energy}. The contour plots depict the local energy
density $\FC_2(\kappa_1,\kappa_2)$ for a fixed value of $\mu_1=\mu_2$
as function of the two Fermi momenta $\kappa_1$ and $\kappa_2$ for two
different positive s-wave scattering lengths $a_0$. The left panel
($a_0/\ell=0.07$) depicts a standard case where the energy density has
a single minimum at $\kappa_1 = \kappa_2 \ne 0$. This class of
solutions with identical density profiles for both components was
already discussed in the context of collapse due to attractive
interactions.

A new class of solutions appears, if the strength of the s-wave
repulsion is increased further. As shown in the right panel of figure
\ref{fig:sep_2comp_energy} ($a_0/\ell=0.12$) the minimum at equal
Fermi momenta has evolved into a saddle point and two energetically
degenerate minima emerge at $\kappa_1=0, \kappa_2>0$ and $\kappa_1>0,
\kappa_2=0$, respectively. For these interaction strengths the density
of one component is depleted or even zero and the density of the other
one is increased in a central region of the trap.

\subsection{Stability conditions}
\label{sec:sep_2comp_stabcond}

The onset of component demixing is indicated by the transition of
the minimum at $\kappa_1=\kappa_2$ into a saddle point. We can derive
a relation that characterizes the onset of component separation by
looking at the determinant of the Hesse second derivative matrix
\eq{ \label{eq:sep_2comp_dethesse}
  D(\rho_1,\rho_2) = \left|\begin{array}{c c}
    \frac{\partial^2\FC_2}{\partial\rho_1^2} &
    \frac{\partial^2\FC_2}{\partial\rho_1 \partial\rho_2} \\
    \frac{\partial^2\FC_2}{\partial\rho_2\partial\rho_1} &
    \frac{\partial^2\FC_2}{\partial\rho_2^2}
  \end{array} \right|  ,
}
where in $\FC_2[\kappa_1(\rho_1),\kappa_2(\rho_2)]$ the Fermi momenta
are expressed in terms of the densities. Again, it is convenient to
use derivatives with respect to the densities, since this eliminates
the dependence of the transformed energy on the chemical potential
automatically. Those stationary points of the energy density that are
minima (or maxima) yield $D(\rho_1,\rho_2)>0$ whereas saddle points
are characterized by $D(\rho_1,\rho_2)<0$. The parameter sets where the
transition from a minimum at $\rho=\rho_1=\rho_2$ to a saddle point
happens are determined by $D(\rho,\rho)=0$. This is a straightforward
generalization of the procedure used in section \ref{sec:col_2comp} to
obtain the stability condition for mean-field collapse. Inserting the
energy density \eqref{eq:sep_2comp_transf_energydens} into
\eqref{eq:sep_2comp_dethesse} leads to
\eq{
  [2 a_0 \kappa + 4 (a_1 \kappa)^3 + \pi]
  [-2 a_0 \kappa + 2 (a_1 \kappa)^3 + \pi] = 0 .
}
The first factor resembles the stability condition
\eqref{eq:col_2comp_stabilitycond_kF} for mean-field collapse of the
Fermi gas. The second factor gives a stability condition for component
demixing. The configuration of overlapping identical density
profiles of the two-components is stable only if the condition
\eq{ \label{eq:sep_2comp_stabilitycond_kF}
  a_0\,\kappa(\xV) - [a_1 \kappa(\xV)]^3 \le \frac{\pi}{2}
}
is fulfilled everywhere in the trap. In regions of the trap where this
condition is violated the overlapping configuration is unstable
against demixing of the components.

As in the case of collapse it is convenient to consider the different
combinations of signs of the s- and p-wave scattering lengths
separately:
\begin{description}
\item[$a_0>0, a_1>0$:] with increasing s-wave scattering length the
critical density for component separation decreases. A
repulsive p-wave interaction stabilizes, i.e., it increases the critical
density. If the ratio of p-wave and s-wave scattering length exceeds
the limit
\eq{ \label{eq:sep_2comp_nolimitcond}
  \frac{a_1}{a_0} \ge \frac{2^{4/3}}{3 \pi^{2/3}} \approx 0.392 ,
}
then the overlapping configuration is absolutely stabilized; no
component separation occurs at all.
\item[$a_0>0, a_1<0$:] the attractive p-wave interaction destabilizes,
i.e., reduces the maximum density of the overlapping configuration. If
the p-wave attraction is sufficiently strong such that
\eq{ \label{eq:sep_2comp_sep_before_col}
  \frac{|a_1|}{a_0} > 2\, (3/\pi)^{2/3} \approx 1.939 ,
}
then the critical density for a mean-field collapse is lower than the
critical density for component separation. Therefore no component
separation but a mean-field collapse of the overlapping configuration
occurs.
\item[$a_0\le0$:] component separation does not occur, even if the p-wave
  interaction is strongly repulsive.
\end{description}
The twisted behaviour of the p-wave interaction with respect to
component separation can be understood from the structure of the
energy density \eqref{eq:sep_2comp_transf_energydens}. The driving
force of component separation---the s-wave interaction---contributes a
term proportional to the product of the densities of both
components. Thus for repulsive s-wave interactions a configuration
with fully overlapping density distributions is higher in energy than
a separated configuration with small overlap but higher absolute
densities.  The p-wave interaction, however, has a term
$[\rho_1\,\rho_2^{5/3} + \rho_1^{5/3}\rho_2]$ proportional to the
product of the densities to some power and an additional term that
depends on the sum of the densities, $[\rho_1^{8/3} +
\rho_2^{8/3}]$. Thus a strongly repulsive p-wave interaction
disfavours both, large overlaps of the two density distributions and
large densities of a single component. Therefore a p-wave repulsion
alone cannot induce component separation. Moreover, it stabilizes
against separation in connection with a repulsive s-wave interaction.
\begin{figure}
\begin{indented}\item[]\hskip-0.085\textwidth
\includegraphics[height=0.28\textheight]{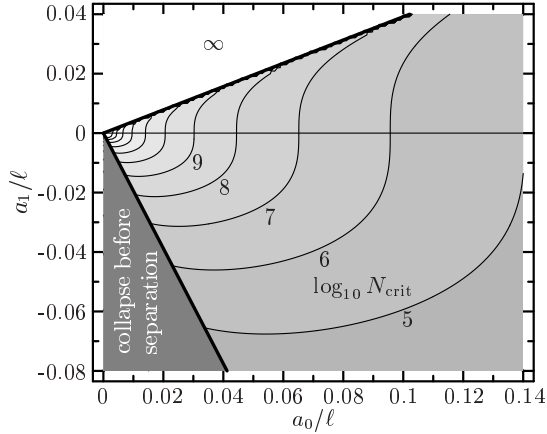}
\end{indented}
\vskip-2.3ex
\caption{Logarithmic contour plot of the critical particle number
  $N_{\mathrm{crit}}$ up to which the trapped two-component Fermi gas
  is stable against component separation as function of the ratio of
  s-wave and p-wave scattering lengths to mean oscillator length
  $\ell$ of the parabolic trap. The white area in the upper part
  indicates the parameter region, where no separation occurs. }
\label{fig:sep_2comp_stabmap}
\end{figure}

A pictorial overview of the influence of s- and p-wave interactions on
the onset of component separation is given by the stability map in
figure \ref{fig:sep_2comp_stabmap}. The contour plot shows the
logarithm of the critical particle number $N_{\mathrm{crit}}$ as
function of the s- and p-wave scattering lengths. This critical
particle number is determined from the solution of the extremum
condition \eqref{eq:ef_extremumcond} that reaches the critical Fermi
momentum \eqref{eq:sep_2comp_stabilitycond_kF} in the trap
centre. Like in section \ref{sec:col_2comp} a deformed parabolic trap
of mean oscillator length $\ell$ is assumed. As soon as the particle
number $N=N_1=N_2$ exceeds the critical particle number
$N_{\mathrm{crit}}$ a demixing of the components appears in the trap
centre. The plot clearly reveals the influence of the p-wave
interaction: Attractive p-wave interactions, on the one hand, reduce
the critical particle number. Moreover, for those interaction
parameters in the dark gray area the p-wave attraction is so strong
that the mean-field collapse of the overlapping configuration happens
before the critical density for component separation is
reached. Repulsive p-wave interactions, on the other hand, stabilize
the overlapping configuration against component separation and prevent
separation completely for interaction parameters in the white region
in figure \ref{fig:sep_2comp_stabmap}.

\subsection{Structure of the separated configurations}

With a few examples we want to discuss typical shapes of the
demixed density distributions that appear beyond the critical
particle number. To discuss the general phenomenology we restrict
ourselves to pure s-wave interactions ($a_1/\ell=0$) and equal
particle numbers $N=N_1=N_2$ in a spherical symmetric trapping
potential.

Even for fixed particle numbers and interaction strengths there exists
a rich variety of different density distributions that may appear in
case of component separation. In general it depends on the detailed
dynamics of the system which of the energetically (almost) degenerate
states is realized. We can distinguish two major classes of separated
solutions: Those with equal chemical potentials $\mu_1=\mu_2$ for both
components and those with $\mu_1\ne\mu_2$.

\begin{figure}
\begin{indented}\item[]\hskip-0.08\textwidth
\includegraphics[width=0.85\textwidth]{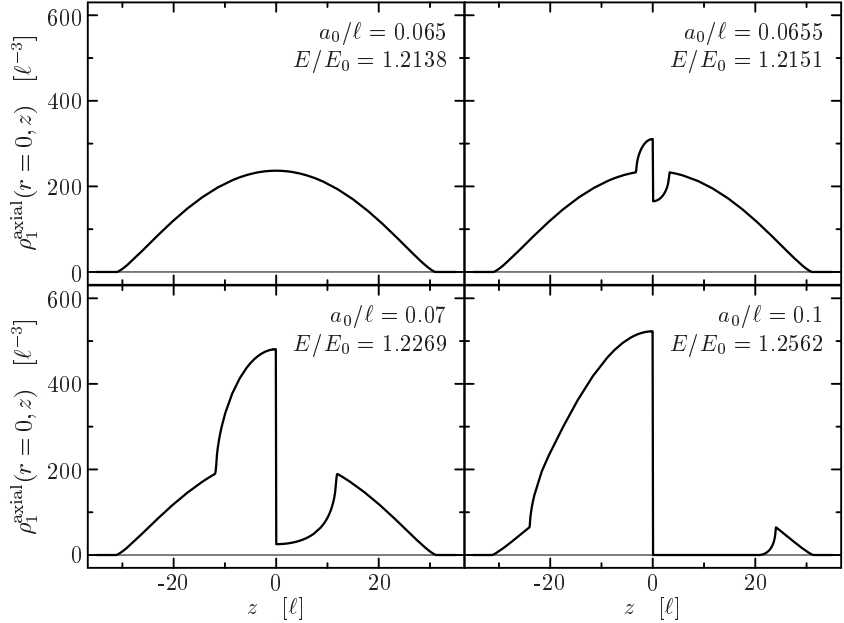}
\end{indented}
\vskip-2.3ex
\caption{Axially symmetric configuration: density profile 
$\rho_1^{\mathrm{axial}}(r=0,z)$ along the $z$-axis for two-component
Fermi gases with $N_1=N_2=10^7$ and different s-wave scattering
lengths $a_0/\ell$. The distribution of the second component
$\rho_2^{\mathrm{axial}}(r=0,z) = \rho_1^{\mathrm{axial}}(r=0,-z)$ is
obtained by inversion of the ordinate.  \emph{An animated GIF of the
density distributions is available from the article's abstract page in
the online journal \cite{TrapWeb}.}}
\label{fig:sep_density_profiles_cut}
\end{figure}

For equal chemical potentials $\mu_1=\mu_2$ [more precisely
$\mu_1-U_1(\xV)=\mu_2-U_2(\xV)$] the extremum condition
\eqref{eq:ef_extremumcond} is symmetric with respect to exchange of
the two components. If the s-wave scattering length is large enough
for phase demixing then there are two energetically degenerate
solutions $\{\rho_1^{(A)}(\xV), \rho_2^{(A)}(\xV)\}$ and
$\{\rho_1^{(B)}(\xV), \rho_2^{(B)}(\xV)\}$ (compare figure
\ref{fig:sep_2comp_energy}), where
\eq{ \label{eq:sep_exchange_symmetry}
  \rho_1^{\mathrm{(A)}}(\xV) = \rho_2^{\mathrm{(B)}}(\xV)
  ,\qquad   
  \rho_2^{\mathrm{(A)}}(\xV) = \rho_1^{\mathrm{(B)}}(\xV) .
}
In general to construct density profiles with equal number of
particles in both components we can switch between the two
solutions depending on the position in the trap. The most simple
ansatz is to take solution (A) in one hemisphere and (B) in the other
\eq{ \label{eq:sep_density_profiles_cut}
  \rho_{\xi}^{\mathrm{axial}}(r,z)
  = \cases{
    \rho_{\xi}^{\mathrm{(A)}}\big(\sqrt{r^2+z^2}\big) & for $z\ge0$ \\
    \rho_{\xi}^{\mathrm{(B)}}\big(\sqrt{r^2+z^2}\big) & for $z<0$
  }
}
which guarantees $N_1=N_2$. Examples for the density profiles
$\rho_1^{\mathrm{axial}}(r,z)$ along the $z$-axis obtained for
different s-wave scattering lengths are shown in figure
\ref{fig:sep_density_profiles_cut}. The upper left panel corresponds
to an interaction strength that is slightly to weak to cause
separation. A tiny increase of $a_0/\ell$ leads to the onset of
component separation (upper right panel), a small density wiggle
appears on top of the smooth density profile in the trap
centre. Further increase of the s-wave repulsion enhances this
demixing and finally leads to a nearly complete spatial separation
where one component is localized in one hemisphere and the second one
in the other hemisphere (lower right panel).

The ansatz \eqref{eq:sep_density_profiles_cut} is by no means
unique. One could subdivide the trap volume into several domains and
switch between the two solutions in each of them. For example, in a
spherical trap one could imagine domains that look like pieces of a
cutted pie. Within the Thomas-Fermi approximation the total energy is
not affected. However, if we go beyond the Thomas-Fermi approximation,
e.g. by including gradient corrections to the kinetic energy, the strong
density variations at the interface of the domains produce a positive
energy contribution. For typical particle numbers of $10^6$ the
relative contribution of the gradient corrections for a cutted
distribution like \eqref{eq:sep_density_profiles_cut} is in the order
of $10^{-5}$ \cite{Roth00}. Even though this is negligible, we assume
that the configuration with minimal interface is preferred---the
axially symmetric profile \eqref{eq:sep_density_profiles_cut} in this
case.
\begin{figure}
\begin{indented}\item[]\hskip-0.08\textwidth
\includegraphics[width=0.85\textwidth]{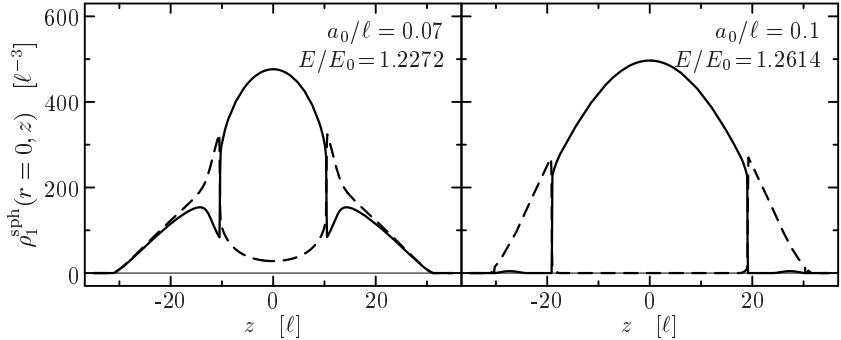}
\end{indented}
\vskip-2.3ex
\caption{Spherical configuration: density profiles 
  $\rho_{1}^{\mathrm{sph}}(r=0,z)$ (solid line) and
  $\rho_{2}^{\mathrm{sph}}(r=0,z)$ (dashed line) along the $z$-axis with
  $N_1=N_2=10^7$ for different s-wave scattering lengths $a_0/\ell$.}
\label{fig:sep_density_profiles_sph}
\end{figure}

There is also a class of spherically symmetric solutions with $N_1=N_2$
that was discussed by several authors \cite{SaPo00,AmMe00}. However,
it turns out that these spherical solutions are higher in energy than
the axial ones. In figure \ref{fig:sep_density_profiles_sph} two
examples for these spherical configurations are shown. The plots also
contain the respective values of the total energy in units of the
energy $E_0=\frac{3N}{2m\ell^2} \sqrt[3]{6N}$ of the noninteracting
two-component gas with the same particle number. For comparison the
energies of the axially symmetric configurations are given in figure
\ref{fig:sep_density_profiles_cut}. It should be noted that in the
spherical case the components have slightly different chemical
potentials.

In an equilibrium situation or during a quasi-stationary process that
leads from overlapping density distributions $\rho_1(\xV)=\rho_2(\xV)$ to a
separated configuration, e.g. by slowly increasing the s-wave scattering
length using a Feshbach resonance \cite{Bohn00}, we expect the
energetically lowest, i.e. the axial configuration with least
interface shown in figure \ref{fig:sep_density_profiles_cut}, to be
realized. This evolution of the density distributions is illustrated by
an animation that is available from the article's abstract page in the
online journal or from \cite{TrapWeb}.

\section{Summary: Phase diagram of trapped degenerate Fermi gases}
\label{sec:phasediag}
\begin{figure}
\includegraphics[width=\textwidth]{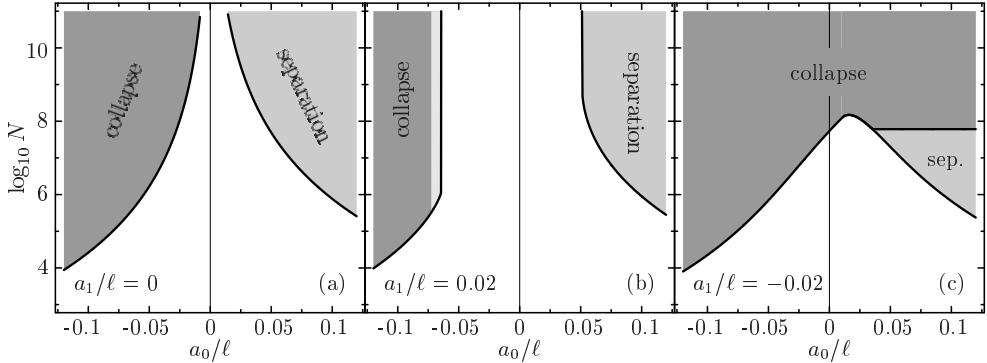}
\vskip-1.3ex
\caption{Zero-temperature phase diagrams of a two-component Fermi
gas in the parameter plane spanned by s-wave scattering length
$a_0/\ell$ and particle number $N=N_1=N_2$. The three diagrams
correspond to different values of the p-wave scattering length: (a)
$a_1/\ell=0$, (b) $a_1/\ell=0.02$, (c) $a_1/\ell=-0.02$. The white
area marks the parameter region where both components show identical
overlapping density profiles. The dark gray area indicates the region
of mean-field collapse and the light gray region characterizes
parameter combinations where both components separate (see text).
\emph{An animated GIF of the phase diagram is available 
from the article's abstract page in the online journal \cite{TrapWeb}.}}
\label{fig:phasediag_2comp}
\end{figure}

To summarize our findings on the stability against mean-field collapse
and component separation we construct a phase diagram of the
two-component Fermi gas at zero temperature. The relevant parameters
are the s- and p-wave scattering lengths and the particle number
$N=N_1=N_2$. Figure \ref{fig:phasediag_2comp} shows the phase diagram
in the plane spanned by the s-wave scattering length $a_0/\ell$ and
the logarithm of the particle number $N$ for three different values of
the p-wave scattering length $a_1/\ell$. We can identify three classes
of density distributions---or phases---of the trapped two-component
Fermi gas: identical overlapping density distributions for both
components (white region), partially or completely demixed density
profiles (gray region), and the p-wave stabilized high-density phase
(light gray area). In addition the collapse of the system due to an
attractive mean-field can occur (dark gray).

In absence of the p-wave interaction the phase diagram has a rather
simple structure shown in part (a) of figure
\ref{fig:phasediag_2comp}. If the s-wave interaction is repulsive
($a_0>0$) the system enters the separated phase if the quantity
$\sqrt[6]{N} a_0/\ell$ exceeds a limiting value. Similarly the gas
collapses in the presence of attractive s-wave interactions if the
quantity $\sqrt[6]{N} a_0/\ell$ exceeds a another maximum value.

The inclusion of p-wave interactions modifies this picture
significantly. Panel (b) of figure \ref{fig:phasediag_2comp} depicts
the phase diagram for a rather weak repulsive p-wave interaction with
$a_1/\ell=0.02$. For s-wave interaction strengths below a certain
limit the overlapping phase extents up to very high particle numbers;
the p-wave repulsion generates absolute stabilization against collapse
and component separation. Moreover, the p-wave stabilized high density
phase emerges near the threshold for absolute stabilization against
collapse (light gray region).

Dramatic effects on the phase diagram are caused also by weak
attractive p-wave interactions as shown in part (c). The maximum
particle numbers possible in the overlapping phase are significantly
reduced, i.e. the p-wave attraction promotes collapse as well as
component separation. In addition it causes the collapse of the
separated configuration, an effect that is in close relation to the
collapse of a single component Fermi gas. If the density or Fermi
momentum of one of the components in the separated configuration
exceeds the critical value given by \eqref{eq:col_1comp_stabilitycond}
then this component will collapse. Actually this causes a distinct
reduction of the parameter region where a stable separated
configuration occurs.

Our investigations show that the p-wave interaction can have big
influence on the properties of trapped degenerate Fermi gases. In the
single-component system it is the origin of the mean-field
instability. In a two-component systems it modifies the phase diagram
substantially and gives rise to completely new phenomena like the
absolute stabilization against s-wave induced collapse and
separation. In this way the p-wave interaction may be very useful for
the envisioned transition to a superfluid state in trapped dilute
Fermi gases.

\section*{References}


\end{document}